# Optimising data processing for nanodiamond based relaxometry

Thea A. Vedelaar[1], Thamir H. Hamoh[1], Felipe P. Perona Martinez[1], Mayeul Chipaux[*,2], Romana Schirhagl[*,1]

1 Groningen University, University Medical Center Groningen, Antonius Deusinglaan 1, 9713 AW.

2 Institute of Physics, École Polytechnique Fédérale de Lausanne (EPFL), CH-1015 Lausanne, Switzerland

* *mayeul.chipaux@epfl.ch ; romana.schirhagl@gmail.com*

## Abstract

The nitrogen-vacancy (NV) center in diamond is a powerful and versatile quantum sensor for diverse quantities. In particular, relaxometry (or T1), allows to detect magnetic noise at the nanoscale. While increasing the number of NV centers in a nanodiamond allows to collect more signal, a standardized method to extract information from relaxometry experiments of such NV ensembles is still missing. In this article, we use T1 relaxation curves acquired at different concentrations of gadolinium ions to calibrate and optimize the entire data processing flow, from the acquired raw data to the extracted T1. In particular, we use a bootstrap to derive a signal to noise ratio (SNR) that can be quantitatively compared from one method to another. At first, T1 curves are extracted from photoluminescence pulses. We compare integrating their signal through an optimized window as performed conventionally, to fitting a known function on it. Fitting the decaying T1 curves allows to obtain the relevant T1 value. We compared here the three most commonly used fit models that are, single, bi, and stretched-exponential. We finally investigated the effect of the bootstrap itself on the precision of the result as well as the use of a rolling window to allows time-resolution.

## Introduction

Diamond based quantum sensing methods have gained attention over the last few years. Through Optically Detected Magnetic Resonance (ODMR)[1] the Nitrogen-Vacancy defect in diamond, in its negatively charged state (NV-), differing from it neutral state (NV$^0$) possesses fluorescence properties which depend on the surrounding magnetic field [2]. Working at room temperature [2,3] and allowing for nanoscale resolution [4–7] the NV- center (hereafter just noted NV center) offers many tremendous experimental assets [8] as well as noteworthy sensitivities [9–11] for sensing variables like temperature [12,13], pH [14,15], strain [16], electric [17] or magnetic fields [4,18], microwave signals [19–21] and electrical current [22].

Among the NV based sensing methods, T1 relaxometry enables sensing magnetic noise without any microwave excitation [23]. As described in Figure 1(a), a pulse of light initializes the NV center's electron spin into its the bright ($m_s = 0$) state. Another pulse, applied after different delays (or dark times, τ), probes the proportion of NV centers which have returned to a darker equilibrium between $m_s = 0$, +1 or −1 states. The time this process takes is called relaxation time or T1 is shortened by magnetic noise. This was initially observed in pioneering works[23–27] and further quantified in [28]. The method was originally used to detect spin labels such as gadolinium ions [23,25,27] or to monitor chemical reactions [15,29,30]. In our group, T1 relaxometry has turned out to be particularly useful for measuring free-radicals (chemicals with unpaired electrons) that are generated by the metabolism of individual cells [31,32].

The relaxometry curve for a single NV center is well understood and can be fitted by a single exponential model [28,33,34]. Using a larger amount of NV centers allows to significantly increase the photoluminescent signal. However, the summation of the signals from different NV centers in varying magnetic environment renders the situation more complex. While few fitting models like the single exponential [24,35], bi-exponential [26,30–32] and the stretched exponential [36,37] decays are often use to analyze such data, there is no clear consensus on which is to be used under different circumstances.

In this work, we use the acquisition procedure developed in [30] with predefined laser intensity and pulse duration. Applying it on a set of 8 nanodiamonds submitted to different gadolinium concentrations, we acquired a calibration data set that we used to systematically compare different ways to extract data from to optimize the whole data processing flow. In particular, we investigate how to best extract the data from the raw photoluminescence pulse train. We then quantitatively compare the three fitting models. To that end, we use a bootstrap approach to obtain a SNR that can be directly compared between the different methods. We also investigate the effect of the bootstrap itself on the result precision as well as the use of a rolling window to obtain temporal information.

1) Materials and methods
    **2.1 Experimental details**

    *a. Sample*

We use gadolinium ions (in the form of $GdCl_3$ in solution) which is a common contrast agent in MRI for magnetic noise to lower T1 in a controlled manner. We use oxygen terminated nanodiamonds with hydrodynamic diameter of 70 nm (Adamas Nanotechnology), identical to the ones used in our intracellular experiments [31,32]. These nanodiamonds are produced by the manufacturer by high pressure high temperature synthesis. Then particles are irradiated with 3MeV electrons at a fluence of $5 \times 10^{19}$ e/cm$^2$ and annealed at the temperature exceeding 700°C. According to the manufacturer, they contain around 500 NV centers on average. To perform a measurement, we first allow the nanodiamonds to attach to a petri dish cover glass before filling it with water. Gadolinium (Gd) solution is then added such that the concentration is increased step wise from 1nM up to 100mM.

    *b. T1 protocol*

One diamond nanoparticle is first identified with our homemade confocal microscope (fully described in [31]). A set of T1 measurements is acquired on that same nanoparticle for each $Gd^{3+}$ concentration. Our previously developed T1 measurement protocol [30] (see Fig 1(a)) consists of a train of 51 laser pulses (532 nm, of (8 mW/µm$^2$)) lasting 5 µ$s$ each, intermitted by 50 dark times exponentially varied from 200 ns to 10 ms. During that sequence, the times at which each photon is detected is stored. We call one execution of a whole sequence of $N_{pulses} = 51$ pulses an observation.

Importantly, depending various parameters such as the nitrogen concentration in the diamond [38], the surface chemistry, the NV center depth [39] and the laser wavelength [40–42] or intensity [43], charge transfer may occur between the negatively charged NV$^-$ and the neutral one NV$^0$ [44,45] . This may significantly impact the relaxation curves obtain as above[46]. However, this phenomenon is mainly relevant in small nanodiamonds with a small number of NV centers or high laser powers and plays a negligible role here.

In previous a work [47], such as proposed in [24], we compared measurements with or without a microwave pulse at resonance with the spin transition and performed the difference to exclude spin impendent processes. In particular, we found that the longer decay temporal constants (bi-exponential fit) and its dependency to the external magnetic noise is preserved. Such acquisitions significantly decrease the contrast of the measurements. We assume in the following that a spin relaxation process is dominating the relaxation curves.

*c. Averaging*

In order to extract sufficient statistics, a measurement sequence (Fig. 1(b)) consists of $N_{obs} = 10^4$ observations acquired on the same particle at each concentration. Those observations are then aggregated either altogether (Fig. 1(b)), through a bootstrap (Fig. 2(a)) or a rolling window aggregation (Fig. 2(b)) (See details in Sec. 2.4). A photoluminescence pulse train is obtained by counting, how many photons are detected at each instant of the pulse sequence (Fig. 1(a)) over all the aggregated observations. Each of the photoluminescence pulses of the train is extracted and isolated such as presented in (Fig. 1(c)) [26]. As described in Sec. 2.2 a T1 relaxation curve (Fig. 1(d)) is extracted from the first microsecond of each pulse as a function of the preceding dark time. The relevant T1 is obtained by fitting a decaying curve (see Sec. 2.3).

This whole process has been reproduced independently on $N_{part} = 8$ different nanoparticles, each submitted to all the investigated $Gd^{3+}$ concentrations.

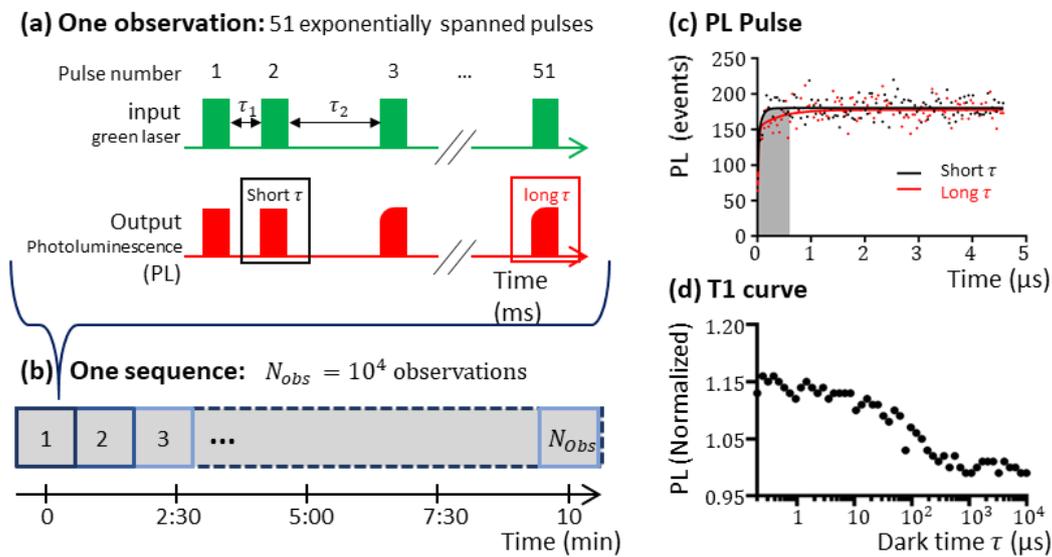

*Figure 1 Principle of $T_1$ measurements:* (a) An observation consists of applying a train of 51 laser pulses (532 nm, of (8 mW/µm²)) of 5 µs each, used both to initialize the and readout the NV center spin state. The 50 darktimes τ between them are exponentially spanned from 0.2 up to $10^4$ µs. Each received photon is timestamped with respect to the laser pulse. (b) each measurement sequence is made of $N_{acq} = 10^4$ successive observations (c) The pulses starting from the second onward are used to determine the T1 by summing the total number of photons received at each timepoint either through the whole $N_{Acq}$ or through the bootstrap or rolling window methods. The black dots indicate the pulse succeeding a dark time of 0.4 µs and the red dots a dark time of 6.8 µs. The intensities of the pulses are integrated over the read window or through pulse fitting. (d) The T1 relaxation curve is obtained either by integrating the first 0.6 µs of each pulse, or through fitting Eq. 1 on it. The photoluminescence is normalized to 1 by dividing by $I_\infty$ obtained from the fit of the Eq. 2, 3 or 4.

## 2.2 Pulse reading

Given the laser power density we apply on the sample, the photoluminescence pulses typically differ only within the few first microseconds.

The photoluminescence signal of a T1 curve can be obtained by integrating the photoluminescence pulses over an optimized window. We use here the first 0.6 µs as previously determined in [30] where this method was used to measure ions in solution and determine the effect from a protein corona.

Alternatively, the photoluminescence can be fitted with eq 1, adapted from [48], which models an ensemble of NV centers.

$$I(t) = A_1(1 - e^{-k_1 t}) + A_2(1 - e^{-k_2 t}) \qquad (1)$$

Here $t$ is the time after the beginning of the pulse. $A_1$ and $A_2$ are related to the different populations in the $m_s = 0$ and $m_s = \pm 1$ states and $k_1$ and $k_2$ are parameters that include but are not limited to the laser power and relaxation rates for the different spin states. To obtain a directly comparable T1 curve, the obtained fitted function is integrating the first 0.6 µs.

### 2.3 Extraction of T1 dynamics

Information about the dynamics of the spin relaxation is obtained from the relaxation curve in the form of a T1 time resulting from a fit. We compared different commonly used fit models including single exponential [34,35,43], bi-exponential [26,33,36] and stretched exponential [36,37,49]. In any case, $\tau$ is the dark time between pulses.

The simplest investigated model is a single exponential decay (Eq. 2). Applied on an ensemble, it computes a global relaxation time of the average of all NV centers.

$$I(t) = I_\infty \left(1 + C_1 e^{-\tau/T_1}\right) \tag{2}$$

$I_\infty$ is the photoluminescence intensity of final thermal equilibrium at long dark times $\tau$ and $C_1$ is contrast of the relaxation curves.

As empirically found [30], the T1 relaxation curves are often well fitted by a bi-exponential model (Eq. 3) comprising a short $T_S$ and a long $T_L$ component. The shorter is in the range of a few microseconds, the longer is of about a few hundreds of microseconds.

$$I(\tau) = I_\infty \left(1 + C_S e^{-\tau/T_S} + C_L e^{-\tau/T_L}\right) \tag{3}$$

The final model is the stretched exponential in which an additional exponent ($\beta$) ranging from 0 to 1[50] (Eq. 4). This way, the obtained T1 constitutes the global relaxation time of the system.

$$I(\tau) = I_\infty \left(1 + C_2 e^{-(\tau/T_1)^\beta}\right) \tag{4}$$

Lhe potential fluctuactions in laser pulse intensity being averaged as discribed below the T1 relaxation curves are fitted without initial normalization. The samples displayed in Fig. 1D, 3A and 4A are normalized devided by $I_\infty$ for easier comparison.

### 2.4 Data aggregation: Generation of the raw photoluminescence pulse train

All observations defined above can be summed together. For each T1 extraction method (Sec. 2.3), a single T1 value is extracted. Other aggregation methods are also used here.

*a. Bootstrap*

As commonly applied in both medical sciences and signal processing, our data was aggregated using a bootstrap [51]. This method allows to infer the probability density for the fitted parameter [52], derive its maximum likelihood, and a standard deviation (see below).

The principle is to resample the whole data set. A subset of $N_{Sub}$ randomly chosen observations (meaning that each observation can be chosen more than once or not at all) are combined (see Figure 2(a) and **Error! Reference source not found.**). In our case, $N_{sub} = 10^4$ (same as $N_{acq}$). For a given T1 extraction method (Sec. 2.2 and 2.3.) a T1 value is obtained.

To obtain the probability density of $T_1$ this procedure is repeated $N_{boot}$ times. In our case, $N_{boot} = 10^4$ resulting in $10^4$ different values for $T_1$ constituting the probability density using the Kernel Density Estimate (KDE) or Parzen-Rosenblatt window [53]. The kernel was based on the one developed by Botev et al [53] which is based on diffusion equations. The KDE transfers the continuous values into a smooth distribution with a total integral of 1 from which the most likely T1 or contrast values and their confidence interval.

b. Rolling window:

To observe the temporal evolution of T1 over the total duration of acquisition, a rolling window (or rolling average) can be applied. This process, often used in econometric studies[54], is schematically presented in Fig. 3. While we previously used the entire $N_{acq}$ repetitions (which are collected within about 10 min) to compute one T1 value, we here attempt to further divide the measurement to gain time resolution. To perform a rolling window the first T1 was computed for a defined number of repetitions (here 1 to 7000). Afterwards the window was moved by 10 repetitions (this is called shift) and the second T1 was computed for observation 11 to 7010. The window was moved again and this was repeated until the window for the final T1 was computed up to the final repetition. The steps that are needed to perform a rolling window are shown in table 2.

| Table 1 | Bootstrap process | Table 2 | Rolling window process |
|---|---|---|---|
| 1 | Select $N_{sub}$ observations randomly among the $N_{acq}$ allowing to select the same measurement multiple times* | 1 | Select a window size and a shift size |
| 2 | Reconstruct the T1 measurement based on the selected repetitions | 2 | Reconstruct the T1 measurement in the window |
| 3 | Compute the T1 value from the resulting curve | 3 | Compute the T1 for this window |
| 4 | Repeat step 1-4 $N_{boot}$ times to obtain the bootstrap samples | 4 | Move the window by the shift size |
| 5 | Apply kernel density approximation | 5 | Repeat steps 2-4 until the window ends at the last repetition |
| 6 | Compute the applicable statistical properties maximum of likelihood (Signal) and standard deviation (Noise) used in for Fig. 3 and 4** | | |

* the observations are selected randomly, without excluding the ones already selected.
** the signal and noise displayed in Fig. 5 are derived from the statistic over the particles (See sec. 2.5)

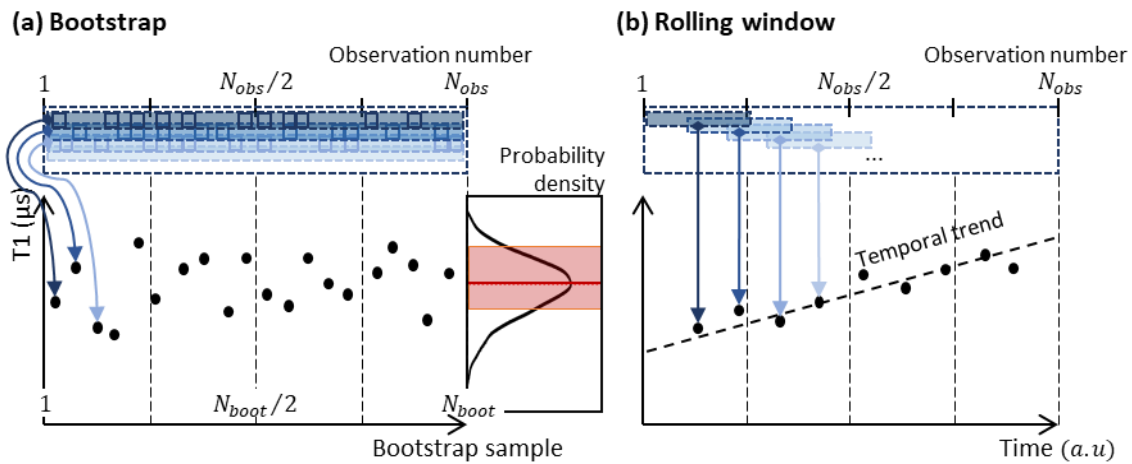

***Figure 2: Data aggregation principle*** *(a) Bootstrap: A subset of the $N_{Sub}$ observations is randomly chosen to generate a T1 relaxation curve from which a T1 value can be fitted. This process is repeated $N_{Boot}$ times generating the same number of T1 values, from which a T1 probability density can be obtained (The most likely value corresponds to the maximum of the curve. The standard). (b) Rolling window: We first generate a T1 from the first $N_{Width}$ observations. We then move the window by $N_{step}$ observations to generate the next T1. This process is repeated until the entire dataset is processed to infer a temporal evolution.*

### 2.5 Signal and Noise errorbar calculation

*a. Statistics over particles*

In case the bootstrap cannot be applied, (Section 3.3, Fig. 5), the signal $S$ is obtained for each concentration by taking the absolute value of the subtraction of the T1 obtained at that concentration to the one in the water condition, averaged over 8 particles. The noise $N$ is taken from the standard deviation over the 8 particles. However, this value includes both the T1 estimation error made by the fit methods and the initial T1 dispersion. Since the second is much larger than the first [30], the ability to discriminate different concentrations is significantly masked by the dispersion of the initial T1 value.

*b. Statistics obtained with bootstrap*

The role of the bootstrap is to place the randomization on the selection of the observation when using a particle rather than on selecting that particle. It therefore allows to derive statistics on the effect of the measurement method itself. In case it can be applied, (Sec. 3.1 and 3.2, Fig. 3 and 4), a signal $S$ is taken from the T1 of maximum of likelihood as obtained by the bootstrap (absolute value of the subtraction to the one of the water condition). The error bars N are also taken from the standard deviation obtained from the bootstrap at the considered concentration. This whole process is repeated and averaged over the 8 different particles.

*c. Signal to noise ratio*

With above definition of $S$ and $N$ the signal to noise ratio is defined as

$$SNR = S/N \quad (5)$$

## 2) Results and Discussions
### 3.1 Pulse fitting

All pulses follow a general pattern shown in (Figure 1(c)). They start with a rapid build-up towards a steady value. The pulses succeeding longer dark times (red) reach the steady value slower than those after shorter ones (black). As described in Section 2.2, the pulses were either integrated over the read window (grey) or fitted with Eq. 1 to better take all the timepoints of the pulse into account and reduce the noise.

The difference between pulses is caused by the probability of the decay from the excited to the ground state via the metastable state which does not emit photons in the detected range [55,56]. When the decay goes through the metastable state, NV center's electrons are also shelved there for a few hundreds of nanosecond before they can cycle again [56,57]. Furthermore, the decay via the metastable state is more likely to occur for electrons in the $m_S = \pm 1$ state [55,56]. These factors create the build-up in the pulses and are the basis of the spin readout of the NV center. A major variable in the polarization model is the excitation rate. Related to the laser intensity, it impacts $k_1$ and $k_2$ in equation 1 directly. While too low pumping does not allow to polarize the NV center fast enough compared to the relaxation, too high energies may ionize the NV⁻ center to NV⁰ [45,58]. In our case, the laser intensity was chosen to be safe for biological samples and kept constant between experiments [31,32].

The fit well captures the build-up to a saturated steady state (ranging up to $\approx 200$ counts). Taking all datapoints into account, the pulse fit should reduce the total relative shot noise. The comparison between the resulting T1 curve obtained from particle 1 exposed to either water or 10 μM, fitted with the bi-exponential model is shown in Figure 3(a).

However, as observed in Figure 3(b) the SNR is not significantly improved when the fit pulse is used.

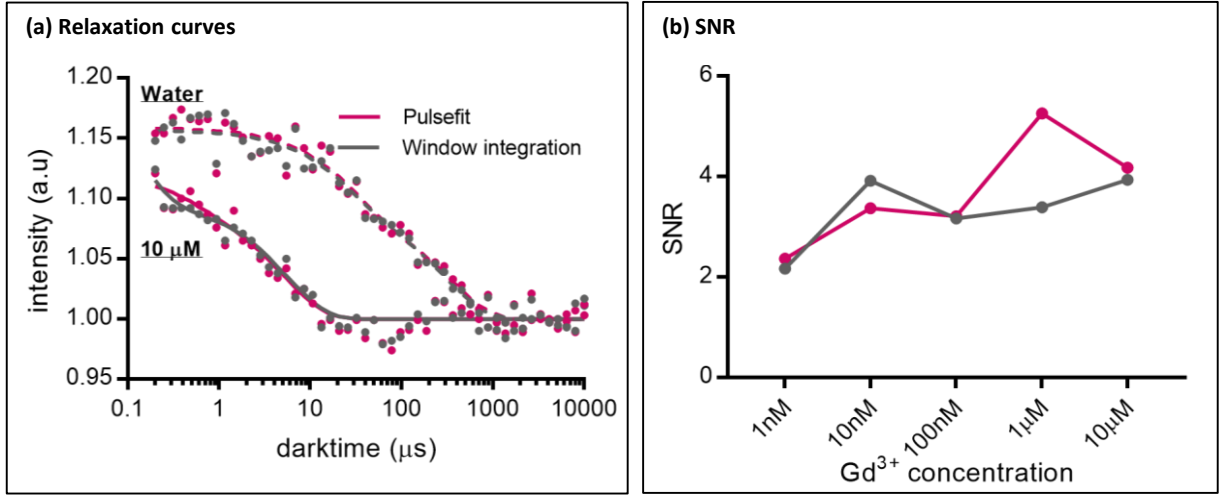

*Figure 3:* *(a) The T1 curves from particle 1 in water and 10 µM $Gd^{3+}$ resulting from window integration (grey) and fitting with Eq. 1 (b) Signal to noise ratio $S/N$ as a function of the $Gd^{3+}$ concentration. The signal and noise are derived from the bootstrap and averaged over 8 particles, as defined in Sec. 2.5.b.*

### 3.2 Exponential fits

We compare the performance of the three different models in differentiating known concentrations of $Gd^{3+}$. Each of these models have their origin in the population dynamics of the NV center. The most important parameters are the difference in intensity between the $m_S = 0$ and $m_S = \pm1$ and the transition rate between these states (expressed in the T1). The single exponential model depicted in Eq. 2 can be naturally derived from the spin relaxation decay dynamics of individual NV center [23,34].

Extending the model to ensembles starts with considering that each NV center has different T1 times. These depend on their orientation and distance to the surface of the nanodiamond [47], the number and proximity of both paramagnetic species and dangling bonds on the surface [33] and nitrogen and $^{13}$C within the diamond. As a result, each NV center feels a slightly different magnetic noise intensity. The relaxation curve obtained from an ensemble is therefore the sum off all those contributions. However, the number of NV centers vary for each nanodiamond and fitting over many variables reduces the precision of the fit and complicates the calculation.

As shown in [50] such a sum over the different T1 can be modeled with a stretched exponential depicted in Eq. 4. This increases the number of fitted parameters to four.

Alternatively, it has been found empirically [30,31,33] that a bi-exponential decay may fit the data well. While under identical conditions (Nanodiamonds, laser pulse intensity and wavelength) the long-time component has been associated to spin relaxation [47], the origin of the short one remains uncertain and may be dominated by charge transfer as discussed in Sec. 2.1. For instance, the laser pumping can induce charge transfer, altering the charge state of the NV centers and the photoluminescence we collect, while the dark time allows it to relax to equilibrium [15,38,44]. Nevertheless, we evidenced in previous work [30] that the longer time constant varies most sensitively with the concentration of samples' than the shorter one.

The fits of these three models on typical T1 relaxation curves are presented in Figure 4(a). It corresponds to the first observed particle exposed to 1 and 100 nM $Gd^{3+}$.

A first measure of the performance of the fits can be read through their residuals which considers the total average squared error made by the fit with respect to the original data. The residuals, averaged over the 8 particles obtained from those fits is presented in the inset of Figure 4(a). At first, it confirms that the bi and stretched exponential decaying models render the situation better than the single exponential one. Despite a clear first shoulder corresponding to the short relaxation decay is often observed subjectively, the stretch and bi exponential's residuals turned out to be quite similar. We could also observe that, weight associated to each decay depends on the nanodiamond, or the Gd concentration. When the last is too high, or for certain particles, only the longer decay persists. In such a case the bi-exponential fit approaches the single or stretched exponential ones.

The degree to which the models explain the difference in T1 as caused by changes in concentration rather than random noise can be measured by computing the (SNR) (see Sec. 2.4). Higher SNR implies that a change in concentration induces better visible difference in the signal with respect to the noise.

Figure 4(b) depicts the most likely T1 values and the standard deviations obtained from the bootstrap at each $Gd^{3+}$ concentrations, both cases averaged over the 8 different particles. Figure 4(c) shows the obtained SNR as defined in Sec. 2.5. For the lower concentrations (1 and 10 nM $Gd^{3+}$), the bi- and stretched-exponentials are more sensitive than the single exponential one. While the bi-exponential has also slightly higher standard deviation, the steepness of the concentration dependency outweighs this disadvantage with respect to the other models.

For higher concentrations however, the error bars linked to the bi-exponential fit increases relatively to the signal. With more fitted parameters, and when the two time constants $T_S$ and $T_L$ get closer, we observe that the fitting procedure may exchange the role of the two, or eventually combine them into a unique exponential decay. While this fit data similarly well, this may render the fit result less predictable.

Nonetheless, our analysis notably confirms that for most of biologically relevant cases i.e. when the T1 is larger than 100 µs, the widely used bi-exponential fit remains well suited.

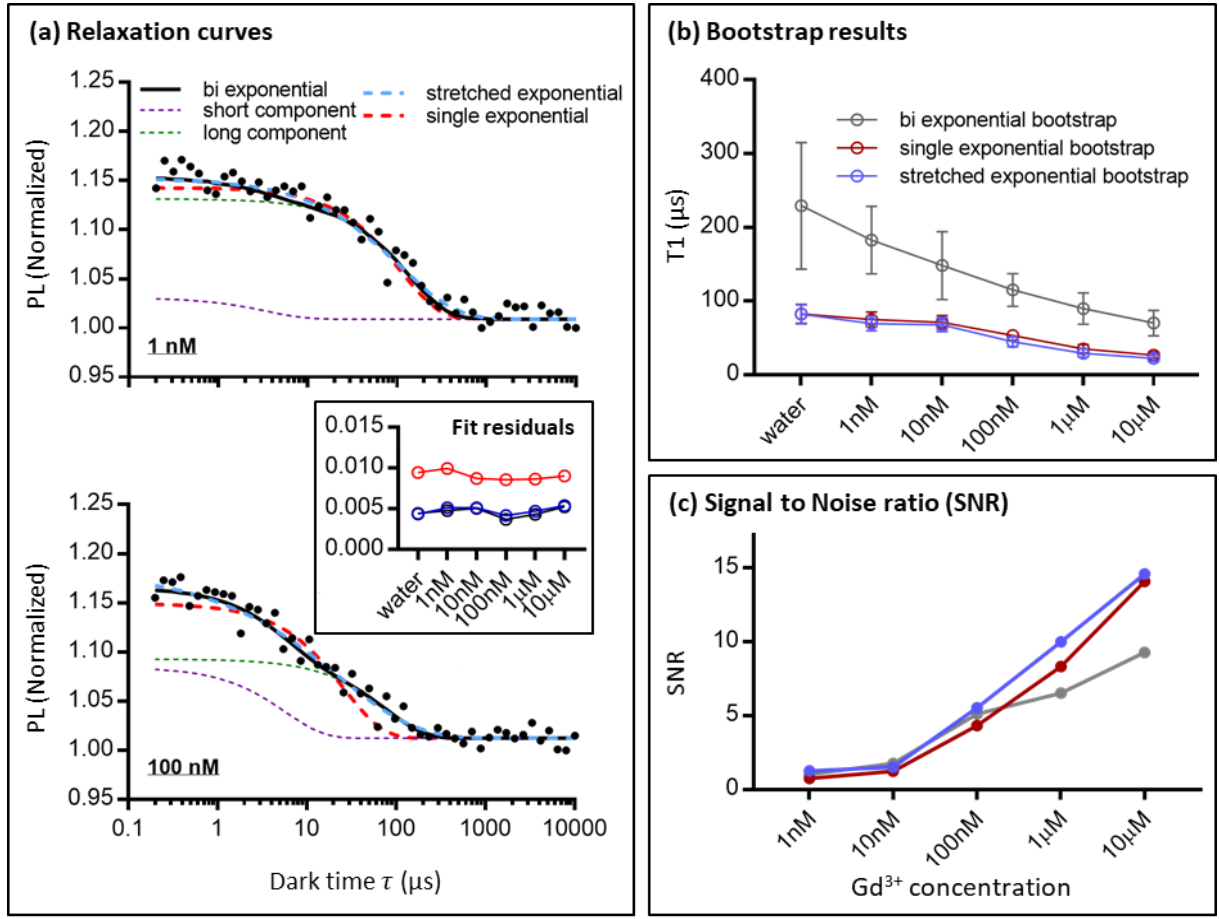

**Figure 4:** (a) Fitted single and bi exponential model for 2 different concentrations of $Gd^{3+}$. $1\ nM$ (top) and $100\ nM$ bottom. The black dots represent the results of the measurement. The figure shows fits for the bi exponential model (solid black), single exponential model (dashed red) and stretched exponential (dashed blue) and the long (green dashed) and short (purple dashed) components of the bi-exponential fits. (b) T1 values as a function of the $Gd^{3+}$ concentration. The signal S and error bars N, are derived from the bootstrap, averaged over the 8 different particles, according to Sec. 2.5.b. (c) Signal to noise ratio $S/N$ as a function of the $Gd^{3+}$ concentration.

### 3.3 Bootstrap

While the bootstrap can be used to compare the different fit model, it can also be applied on the raw data to determine the most likely T1. The idea behind is that some outlier observations may not to be considered in a random manner such that the obtained most likely value can be more robust than the direct fit. Since the bootstrap cannot be used on data that has already processed with a bootstrap, we compare in Figure 5 the fit results taken over all observation to the most likely value obtained from the bootstrap, for both the stretched and bi exponential model. The uncertainty is obtained each time from the standard deviation observed over the 8 different particles at the considered $Gd^{3+}$ concentration. (See Sec. 2.5)

While the T1 values remain quite similar when using the bootstrap, the method did not lead to any improvement. First, it is to be noted that the standard deviations are taken from the 8 different particles. They are therefore significantly enlarged by the initial T1 dispersion. Furthermore, the bootstrap selects observations independently from one another. This both allows to take each observation more than once but also not to take some observations at all. Although both the direct fit and bootstrap take the same amount of observations in total $(N_{Sub} = N_{Acq} = 10^4)$, the number of them that are independent from one another is lower in the bootstrap than in the direct fit. This therefore reduces the averaging and increases the relative noise in the T1 curve and so on the fitted T1. It appears that this is not compensated by allowing potential outlier observations to be ignored.

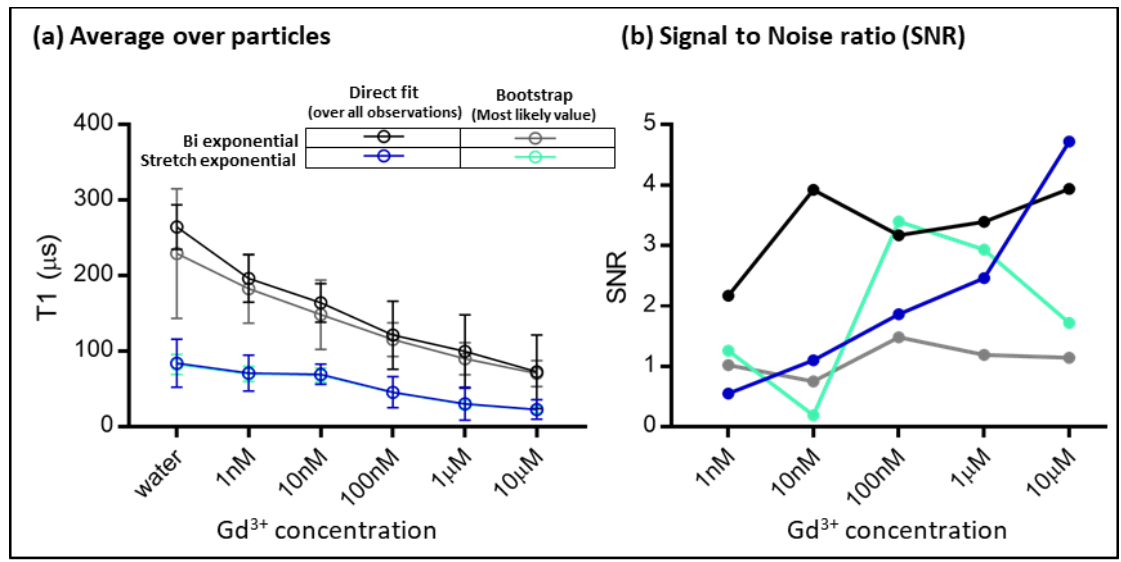

**_Figure 5:_** *(a) T1 values obtained by fitting either a bi exponential or a Stretch exponential model over all the observation at once ("direct fit") or taking the most likely value from the bootstrap. In all cases we averaged over 8 particles. The error bars correspond to the standard deviation of the values obtained above calculated from the data for 8 particles (details in Sec. 2.5.a). (b) Signal to noise ratio as a function of the $Gd^{3+}$ concentration.*

### 3.4 Rolling window

While the previous methods aim to detect a concentration that is constant within one measurement, we here attempt to improve time resolution. To this end we created a measurement consisting of three $Gd^{3+}$ concentrations: water (0nM), 10nM and 1µM (1000nM). This means that the T1 should be high in the beginning of the measurement and low at the end. The rolling window (shown in Fig. 6) used 7000 repetitions with a shift of 10.

This window was intentionally moved over the different concentrations. This simulates a case where we do not know when a concentration change occurs. The points in the purple and green areas of Fig. 6 are composed of data from two different concentrations. The T1 value in each point can be predicted by using the average T1 value of the rolling window of each measurement with a size of 7000. The concentration prediction is then made by calculating the proportion of each measurement in each point and computing the average T1 based on these proportions. The predicted values are shown in Fig. 6 with the dotted lines with different colors representing the different models (black: bi-exponential, red: single exponential, blue: stretched exponential). Fig. 6 shows that all exponential models follow the predicted curve very well.

The most important consideration for choosing the right parameters for a rolling window is the window's size. If the timescale of the expected changes is known this can be used as a guideline for selecting the window size. However, when lowering the window size, we have to consider that each resulting T1 value is then based on a smaller number of repetitions (for an optimization of this parameter see supplementary information).

The rolling window acts as a longpass filter. A too long rolling window would average out changes on short time scales. Oppositely, a too short window will lead to unreliable results. Overall, the window size has to be optimized per set of experiments. In our case, we observed that bellow 7000 observations, the fits are becoming less stable, which reduces the accuracy of the fit.

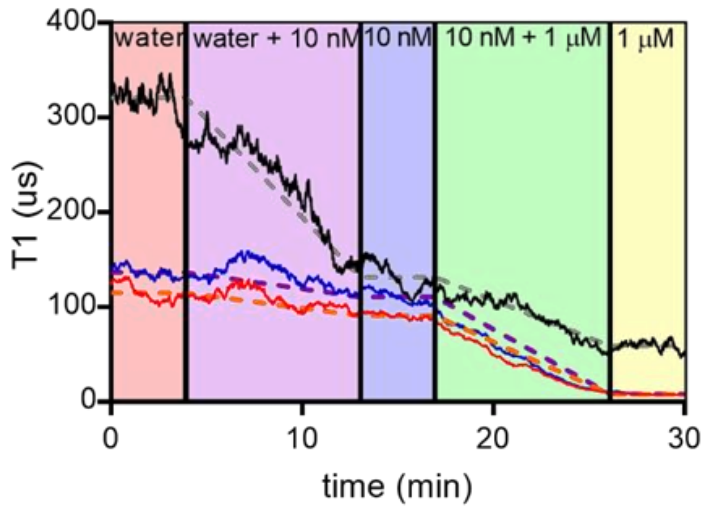

*Figure 6:* The rolling window of 3 measurements for different concentrations. Each window consisted of 7000 repetitions and are shifted by 10 for each sequential point. The solid lines represent the measured T1 values for each point of the figure for the different models; bi-(black), single (red) and stretched (blue) exponentials. The colored blocks represent where each calculation consisted of which concentration (purple for water and 10nM, green for 10nM and 1µM). The dashed lines represent the T1 value obtained from the average of a rolling window of 7000 and are used to represent the average T1 for one specific concentration.

**Conclusion**

In this paper we compared different methods to analyze T1 data: single, bi- and stretched exponential models. These are the most commonly used models to fit the T1 curve. We presented pulse fitting and bootstrap to remove noise from T1 data. Lastly, we showed the rolling window method for improving the temporal resolution of the measurement. We compared all these methods based on a calibration dataset which uses NV centers in nanodiamond to measure different concentrations of gadolinium. We demonstrated that all models and methods can be applied successfully to this data.

By using a bootstrap, we showed that the stretched and bi-exponential fit models are better in differentiating between concentrations, with a preference for the stretched one at higher $Gd^{3+}$ concentrations. For lower concentrations, the bi-exponential has larger variation between measurements, but is also more sensitive to concentration changes. The T1 resulting from the stretched exponential remains very similar to the single exponential but appears to be more predictable. Furthermore, the fit quality, for the bi and stretched exponentials, are significantly better than for single exponential. However, when selecting the best model for the data, the experimental design should be a leading factor as well. While we investigated here nanodiamond with NV centers ensemble, the single exponential model may still be best for single NV centers.

We also presented two alternative methods to compute the T1 from the measurement. The first is based on modelling the pulses from the pulse train. While the output T1 remain unchanged, the measurement quality is not improved in term of SNR. Thus, it might be useful in datasets of worse quality, or when the optimized integration window is not known.

The second concerns the use of the bootstrap to improve the quality of an output T1. Either masked by initial T1 dispersion, or due to plausible limitations that we identified, we could not observe significant improvement here as well.

Lastly, the rolling window was used to show temporal information on the T1. We showed that, in this case, each model renders the decrease in T1 with increasing concentrations rather well. While the T1 is "noisy" as the rolling window moves, the changes induced by combining different $Gd^{3+}$ concentration is larger.

Data Availability Statement: All data is available in this manuscript or its supplementary materials or available on request from then authors.

nanodiamond probe. *Proc. Natl. Acad. Sci. U. S. A.* **110**, 10894–8 (2013).

Author contributions statements:
FPM has performed the experiments performed in this article. TV has performed the data analysis with the help of TH. RS and MC have supervised the project. RS has acquired the funding and leads the research group. TV has written the first version of the manuscript and all authors have edited and approved the final version.

Competing interests statement:
The authors have no competing interests to declare.

# Optimising data processing for nanodiamond based relaxometry

## Supplementary Information

Thea A. Vedelaar, Thamir H. Hamoh, Felipe Perona Martinez, Mayeul Chipaux, Romana Schirhagl

*Optimization of rolling window*

The rolling window for different window sizes was computed to assess the relative error (RE) of the average for each window and each concentration. For each point in figure 5 we calculated the RE for all measurements and took the average and calculated the standard deviation. Outliers were removed by IQR before computing the RE. This figure shows that there is a relation between the RE and the window size. It shows that the RE decreases with window size. Furthermore, the standard deviation of the RE also decreases with window size. What this implies is that for larger window sizes, the T1 over the full rolling window is more stable than for the smaller values. When comparing the different models presented in figure 5, it is clear that the bi exponential model has a few large outliers, as shown by the large standard deviations for smaller windows. However, after a window size of 4500 repetitions it behaves more like the other models. The single and stretched exponentials show less of these extreme outliers. These models still clearly show a decrease in relative error with larger windows. Furthermore, the figure shows that the single exponential model has the smallest RE while the bi-exponential has the highest RE.

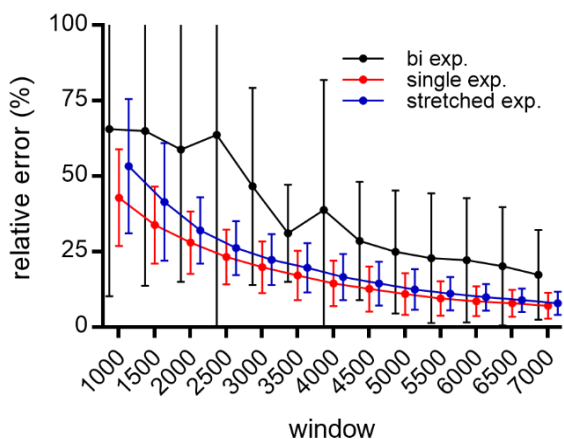

Figure S1: The relative error of the rolling window analysis of all measurements for different window sizes. Each point represents the mean of the relative error for the bi exponential model (black), single exponential (red) and the stretched exponential (blue). The error is computed as the standard deviation of the average of the relative error.